\documentstyle[11pt]{article}
\RequirePackage{comment}
\excludecomment{frontmatter}

\begin{document}

\newcommand{\be}{\begin{equation}}
\newcommand{\ee}{\end{equation}}
\newcommand{\ba}{\begin{array}}
\newcommand{\ea}{\end{array}}
\newcommand{\goto}{\rightarrow}
\newcommand{\longto}{\longrightarrow}
\newcommand{\ov}{\overline}

\begin{frontmatter}

\title{Estimating of $P(Y<X)$in the Exponential case Based on Censored Samples}

\begin{aug}
\author{\fnms{A. M.} \snm{Abd Elfattah}\corref{}\ead[label=e1]{a_afattah@hotmail.com}}
\address{Department of mathematical Statistics, Institute of Statistical Studies and Research\\
Cairo University, Cairo, Egypt}
\end{aug}

\begin{aug}
\author{\fnms{O. Mohamed} \snm{Marwa}}
\address{Department of mathematics, Zagazig University, Cairo, Egypt}
\affiliation{Cairo University and Zagazig University}
\end{aug}

\begin{abstract}
In this article, the estimation of reliability  of a system is discussed $p(y<x)$ when strength, $X$,
and  stress, $Y$, are two  independent exponential distribution with different scale parameters
when the available data are type II Censored sample. Different methods for estimating the reliability
are applied. The point estimators obtained are maximum likelihood estimator, uniformly minimum variance
unbiased estimator, and Bayesian estimators based on conjugate and non informative prior distributions.
A comparison of the estimates obtained is performed. Interval estimators of the reliability are also discussed.
\end{abstract}


\begin{keyword}
\kwd{Maximum likelihood estimator}\kwd{Unbiasedness}\kwd{Consistency}\kwd{Uniform minimum variance unbiased estimator}
\kwd{Bayesian estimator}\kwd{Pivotal quantity}\kwd{Fisher information}
\end{keyword}

\end{frontmatter}

\title{Estimating of $P(Y<X)$in the Exponential case Based on Censored Samples}
\author{Abd Elfattah, A. M.\\
Department of mathematical Statistics, Institute of Statistical Studies and Research\\
Cairo University, Cairo, Egypt.\\\\\\
Marwa O. Mohamed\\
Department of mathematics, Zagazig University, Cairo, Egypt.\\
}

\date{}
\maketitle
\begin{abstract}
In this article, the estimation of reliability  of a system is discussed $p(y<x)$when strength,$X$, and  stress, $Y$ , are two  independent exponential distribution with different scale parameters when the available data are type II Censored sample. Different methods for estimating the reliability are applied . The point estimators obtained are maximum likelihood estimator, uniformly minimum variance unbiased estimator, and Bayesian estimators based on conjugate and non informative prior distributions. A comparison of the estimates obtained is performed. Interval estimators of the reliability are also discussed.
\end{abstract}
{\bf Key Words:} { Maximum likelihood estimator; Unbiasedness; Consistency ; Uniform minimum variance unbiased estimator; Bayesian estimator; Pivotal quantity;  Fisher information .
}
\section*{ 1~Introduction}
$~~~~$In life testing, it is often the case that items drawn from a population are put on test and their times of failure are recorded. For a number of reasons, such as budget or limited time, it is often necessary to terminate the test before all the failure times have been observed; In this case the data become available in some ordered manner. The test is usually terminated after a fixed time or a fixed number of failures are observed giving a censored sample see Epstein and Sobel (1953).
In stress-strength model, the stress  $Y$ and the strength  $X$ are treated as random variables and the reliability of a component during a given period  $(0,T)$is taken to be the probability that the strength exceeds the stress during the entire interval. the reliability  of a component is  $P(Y<X)$.
Several authors have considered different studies for  stress and strength with exponential distribution with complete sample. Tong(1974) discussed the estimation of$P(Y<X)$  in the exponential case.
Tong(1977) had a look at the estimation of $P(Y<X)$ for exponential families. A good review of the literature can be found in Johnson (1988). Beg(1980a,b and c)estimated the exponential family ,two parameter exponential distribution and truncation parameter distributions . Basu(1981) considered maximum likelihood  estimators (MLE) for $P(Y\leq{X})$  in case of gamma and exponential distributions. Sathe and Shah (1981) studied estimation of $P(Y>X)$ for the exponential distribution. Chao (1982) provided simple approximations for bias and mean square error of the maximum likelihood estimators of reliability when stress and strength are independent exponentially distributed random variables.
Awad and Charraf (1986) studied three different estimators for reliability has a bivariate exponential distribution. Dinh, et al (1991) obtained the MVUE of $R$ when  $X$ and $Y$  have the bivariate normal distribution. Moreover, they considered the case when $X$ and $Y$ have the bivariate exponential distribution. Bai and Hong (1992) estimated  $P(Y\leq{X})$ in the exponential case with common location parameter. Kunchur and Mounoli (1993) obtained UMVUE of stress- strength model for multi component survival model based on exponential distribution for parallel system. Siu-Keung Tse and Geoffrey Tso (1996) studied the shrinkage estimation of reliability for exponential distributed lifetime. A note on the UMVUE on $P(Y\leq{X})$ in the exponential case discussed in Cramer and Kamps (1997). Selvavel, et al (2000) studied reliability $R$ when $(X,Y)$ jointly follows a truncated bivariate exponential distribution with a common parameters. Khayar (2001) discussed the reliability of time dependent stress-strength models for exponential and Rayleigh distributions. Shrinkage estimation of $P(Y\leq{X})$ in the exponential case discussed by Ayman and Walid (2003).Tachen (2005)deals with the empirical Bayes testing the reliability of an exponential distribution .

In the present article, the reliability, $R$ , is studied when $X$  and  $Y$  two  independent exponential distribution with different scale parameters. Which can be represented as
$f(x;\alpha)=\frac{1}{\alpha}e^{\frac{-x}{\alpha}},~~~~~~~x>0,\alpha>0.$\\
Different estimators of  $R$  are derived, namely, maximum likelihood estimator (MLE), uniform minimum variance unbiased estimator, (UMVUE), and Bayesian estimators with mean square error loss functions corresponding to conjugate and non informative priors. A comprehensive comparison of the various point estimators (MLE,UMVUE, and Bayes) is performed on the basis of the mean squared error. Interval estimators of $R$ are also discussed. A numerical comparison of the intervals obtained.
\section*{2~Reliability }
Let $X$  be the strength of a component and  $Y$ be the stress acting on it. Let $X$  and $Y$   be exponential independent random variables with parameters $\alpha$  and $\beta$, respectively. That is , the probability density functions (pdfs) of $X$  and $Y$  are, respectively,
$$f(x;\alpha)=\frac{1}{\alpha}e^{\frac{-x}{\alpha}},~~~~~~~x>0,\alpha>0\eqno(2.1),$$
and
$$f(y;\beta)=\frac{1}{\beta}e^{\frac{-y}{\beta}},~~~~~~~y>0,\beta>0\eqno(2.2),$$
where  $\alpha$ and $\beta$  are unknown parameters .\\
The reliability of the component will be\\
$R=P(Y<X)$\\
$~~=\frac{1}{\alpha}\int{0}{\infty}(1-e^{\frac{-x}{\beta}})e^{\frac{-x}{\alpha}}dx$
$$=frac{\alpha}{\alpha+\beta}.\eqno(2.3)$$
if  $\alpha$ and $\beta$ are known then $R$  is simply calculated using Eq.(2.3).
\section*{3.~Point Estimation of $R$ }
\subsection*{3.1~   Maximum Likelihood Estimator of $R$ }
If  $\alpha$ and $\beta$ are unknown the MLE of, $\hat{R_1}$ , of $R$  is given by
$$\hat{R_1}=frac{\hat{\alpha}}{\hat{\alpha}+\hat{\beta}},\eqno(3.1)$$
where $\hat{\alpha}$  and  $\hat{\beta}$ are the MLEs of $\alpha$ and $\beta$ , respectively. For obtaining  $\hat{\alpha}$  and  $\hat{\beta}$  we argue as follows:\\
Suppose that  $r_1$  components where $r_1\leq{n}$  with strengths $X_i$;$i=1,...,r_1$  , each of which having exponential distribution with parameter $\alpha$ as in Eq. (2.1) are subjected, respectively, the likelihood function will be\\
$L=\frac{n!}{(n-r_1)!}\prod_{i=1}{r_1}{\frac{1}{\alpha}e^{\frac{-x_i}{\alpha}}[e^{\frac{-x_{r_1}}{\alpha}}]^{n-r_1}}$
$$L=\frac{n!}{(n-r_1)!}(\frac{1}{\alpha})^{r_1}e^{\frac{-\sum_{i=1}^{r_1}x_i}{\alpha}-\frac{-x_{r_1}(n-r_1)}{\alpha}},eqno(3.2)$$
Taking the logarithm of Eq. (3.2) and find the derivative with respect to $\alpha$ \\
$\frac{dlnL}{\alpha}=\frac{-r_1}{\alpha}+\frac{\sum_{i=1}^{r_1}x_i}{\alpha^2}+\frac{(n-r_1)x_{r_1}}{\alpha^2}=0$\\
$-r_1\alpha=\sum_{i=1}^{r_1}x_i+(n-r_1)x_{r_1},$
$$\hat{\alpha}=\frac{\sum_{i=1}^{r_1}x_i+(n-r_1)x_{r_1}}{r_1},\eqno(3.3)$$
To stress $Y_j$; $j=1,...,r_2$, having exponential distribution with parameter $\beta$  as in Eq. (2.2), where $r_2\leq{m}$  . Assuming that  $X_1$ and $Y_j$;  $i=1,...,r_1$ and $j=1,...,r_2$ , are independent , the likelihood function will be\\
$L=\frac{m!}{(m-r_2)!}\prod_{j=1}{r_2}{\frac{1}{\beta}e^{\frac{-y_j}{\beta}}[e^{\frac{-y_{r_2}}{\beta}}]^{m-r_2}}$
$$L=\frac{m!}{(m-r_2)!}(\frac{1}{\beta})^{r_2}e^{\frac{-\sum_{j=1}^{r_2}y_j}{\beta}-\frac{-y_{r_2}(m-r_2)}{\beta}},\eqno(3.4)$$
Taking the logarithm of Eq. (3.4) and find the derivative with respect to$\beta$\\
$\frac{dlnL}{\beta}=\frac{-r_2}{\beta}+\frac{\sum_{j=1}^{r_2}y_j}{\beta^2}+\frac{(m-r_2)y_{r_2}}{\beta^2}=0$\\
$-r_2\beta=\sum_{j=1}^{r_2}y_j+(m-r_2)y_{r_2},$
The MLE of  $\beta$  will be
$$\hat{\beta}=\frac{\sum_{j=1}^{r_2}y_j+(m-r_2)y_{r_2}}{r_2},\eqno(3.5)$$
Now we shall study some properties of $\hat{R_1}$ .\\
First, if $r_1=r_2=r$ :
\subsection*{1]~Unbiasedness}
$E(\hat{R_1})=\frac{\alpha}{\alpha+\frac{r\beta}{r+1}}[1-\frac{(2r-1)}{r(r-2)}(1-\frac{\alpha}{\alpha+\frac{r\beta}{r+1}})^2]$\\
$\lim_{r\rightarrow{\infty}}E(\hat{R_1})=R-R\lim_{r\rightarrow{\infty}}\frac{(1-R)^2}{(r-1)}$\\
then
$\lim_{r\rightarrow{\infty}}E(\hat{R_1})=R$\\
then, $\hat{R_1}$  asymptotically unbiased estimator of $R$ .
\subsection*{2]~Consistency}
$Var(\hat{R_1})=\frac{(2r-1)}{r(r-2)}[\frac{\frac{r\beta}{(r-1)\alpha}}{\frac{(r+1)\alpha\beta}{(r-1)\beta}}]^2[\frac{1}{1+\frac{r\beta}{(r-1)\alpha}}]^2$\\
$\lim_{r\rightarrow{\infty}}Var(\hat{R_1})=R^2\frac{(\beta}{\alpha}\lim_{r\rightarrow{\infty}}\frac{1}{r}$\\
then\\
$\lim_{r\rightarrow{\infty}}Var(\hat{R_1})=0$\\
then,  $\hat{R_1}$  is a consistent estimator for  $R$.\\
Second, if $r_1\not={r_2}$ :

\subsection*{1]~Unbiasedness}
$E(\hat{R_1})=\frac{\alpha}{\alpha+\frac{r_1\beta}{r_1+1}}[1-\frac{(r_1+r_2-1)}{r_2(r_1-2)}(1-\frac{\alpha}{\alpha+\frac{r_1\beta}{r_1+1}})^2]$\\
For fixed $r_2$,\\
$\lim_{r_1\rightarrow{\infty}}E(\hat{R_1})=\lim_{r_1\rightarrow{\infty}}\frac{\alpha}{\alpha+\frac{r_1\beta}{r_1+1}}[1-\frac{(r_1+r_2-1)}{r_2(r_1-2)}(1-\frac{\alpha}{\alpha+\frac{r_1\beta}{r_1+1}})^2]$\\
$\lim_{r_1\rightarrow{\infty}}E(\hat{R_1})=R[1-\frac{1}{r_2}(1-R)^2]$\\
then
$\lim_{r_1,r_2\rightarrow{\infty}}E(\hat{R_1})=R$\\
then, $\hat{R_1}$  asymptotically unbiased estimator of $R$ .
\subsection*{2]~Consistency}
$Var(\hat{R_1})=\frac{(r_1+r_2-1)}{r_2(r_1-2)}[\frac{\frac{r_1\beta}{(r_2-1)\alpha}}{\frac{(r_1+1)\alpha\beta}{(r_2-1)\beta}}]^2[\frac{1}{1+\frac{r_1\beta}{(r_1-1)\alpha}}]^2$\\
For fixed $r_2$,\\
$\lim_{r_1\rightarrow{\infty}}Var(\hat{R_1})=\lim_{r_1\rightarrow{\infty}}\frac{(r_1+r_2-1)}{r_2(r_1-2)}\lim{r_1\rightarrow{\infty}}[\frac{\frac{r_1\beta}{(r_2-1)\alpha}}{\frac{(r_1+1)\alpha\beta}{(r_2-1)\beta}}]^2\lim_{r_1\rightarrow{\infty}}[\frac{1}{1+\frac{r_1\beta}{(r_1-1)\alpha}}]^2$\\
and,\\
$\lim_{r_1,r_2\rightarrow{\infty}}Var(\hat{R_1})=R^2[\frac{\beta}{\alpha}]^4\lim_{r_2\rightarrow{\infty}}\frac{1}{r_2}$\\
then\\
$\lim_{r_1,r_2\rightarrow{\infty}}Var(\hat{R_1})=0$\\
then,  $\hat{R_1}$  is a consistent estimator for  $R$.
\subsection*{3.2~   Uniform Minimum Variance Unbiased Estimator of $R$ }
Let  $X_1,...,X_{r_1}$and$Y_1,...,Y_{r_2}$  be two independent random samples , of size  $r_1$ and$r_2$ , respectively, drawn from exponential distributions with parameters $\alpha$ and $\beta$ , respectively,\\
Define\\
$z_i=lne^{x_i}$,$v_j=lne^{y_j}$,$i=1,...,r_1$and$j=1,...,r_2$,\\
$Z=\sum{i=1}{r_1}z_i$,and$V=\sum{j=1}{r_2}v_j$\\
Clearly from Eq. (3.2) and (3.4) we see that $Z$,$V$ is a complete sufficient statistic for $\alpha$ , $\beta$.\\
Now, we have\\
$E(W)=1.P(v_1<z_1)+0.P(v_1\geq{Z_1})$\\
$E(W)=P(lne^{y_j}<lne^{x_i})$\\
$E(W)=P(y<x)=R$\\
Let$W$ be the indicator variable  $I_{[0,z_1)}(W)$.it could be seen that $W$ be unbiased estimator for $R$ , by using Rao-Black Well and Lehmann-Scheff$\acute{e}$  we have $\hat{R_2}$  is UMVUE for $R$ .(see Mood et al (1974)).\\
$\hat{R_2}=E(W/Z,V)$\\
$\hat{R_2}=\int_{z_1}\int_{v_1}wf(z_1,v_1/Z,V)dv_1dz_1$\\
where  $f(z_1,v_1/Z,V)$ is the conditional pdf of $z_1$,$v_1$ given $Z$,$V$  .
Notice that $z_1$and $v_1$  are independent exponential random variables with parameters $\alpha$ and $\beta$, respectively, and that  $Z$ and  $V$ are independent gamma random variables with parameters  $(n,\alpha)$ and $(m,\beta)$ , respectively.\\
We see that $Z-z_1$ and $V-v_1$ are impendent gamma random with parameters $(n-1,\alpha)$ and $(m-1,\beta)$   , respectively.  Moreover $Z-z_1$ and $z_1$ ,as well as $V-v_1$  and $v_1$  are also independent. We see that  \\
$\hat{R_2}=\int_{z_1}\int_{v_1}w\frac{\Gamma(r_1)(Z-z_{1})^{(r_1-2)}\Gamma(r_2)(V-v_{1})^{(r_2-2)}}{\Gamma(r_1-1)(z_1)^{(r_1-1)}\Gamma(r_2)(v_1)^{(r_2-1)}}dv_1dz_1$
$$\mbox{\cal $\hat{R_{2}}$}=\frac{\Gamma(r_1)\Gamma(r_2)}{\Gamma(r_1-1)\Gamma(r_2-1)z^{(r_1-1)}v^{(r_2-1)}}\left\{ \begin{array}{cc}
\int_{0}^{v}\int_{v_1}^{z}(V-v_1)^{(r_2-2)}(Z-z_1)^{(r_1-2)}dz_1dv_1, & v_1<{z_1},\\
\int_{0}^{z}\int_{0}^{z_1}(Z-z_1)^{(r_1-2)}(V-v_1)^{(r_2-2)}dv_1dz_1, & v_1\geq{z_1};
\end{array}
\right.\eqno(3.6)$$

The computation of the UMVUE $\hat{R_2}$ is very complicated as it can seen from equation (3.6).so, we will use the MATHCAD program to evaluate the value of $\hat{R_2}$ .
\subsection*{3.3.~Bayes Estimator of $R$ }
We obtain Bayes Estimator of  $R$  with respect to the mean square error loss function with respect to conjugate and non informative prior distributions.

\subsection*{3.3.1. Conjugate gamma prior distribution}
Let $X_1,...,X_{r_1}$and$Y_1,...,Y_{r_2}$  be the first $r_1$ and $r_2$ failure observations from $X_1,...,X_n$ and $Y_1,...,Y_m$ respectively, where both of them have exponential distribution with parameters $\alpha$and$\beta$ respectively. Assume that the prior distribution of $\alpha$ is given by\\
$\pi_{01}=f(\alpha)=\frac{v_1^{u_1}}{\Gamma(u_1)}(\frac{1}{\alpha})^{u_1-1}e^{\frac{-v_1}{\alpha}},u_1,v_1,\alpha>0$\\
the likelihood function with type II censored sample is , respectively,
$$f(x_1,...,x_{r_1}|\alpha)=\frac{n!}{(n-r_1)!}(\frac{1}{\alpha})^{r_1}e^{-\frac{1}{\alpha}(\sum_{i=1}{r_1}x_i+x_{r_1}(n-r_1))},\eqno(3.7)$$
and
$$f(y_1,...,y_{r_2}|\beta)=\frac{m!}{(m-r_2)!}(\frac{1}{\beta})^{r_2}e^{-\frac{1}{\beta}(\sum_{j=1}{r_2}y_j+y_{r_2}(m-r_2))},\eqno(3.8)$$
Assuming that $\alpha$and$\beta$  are independent having prior gamma distributions, the posterior distributions of  $\alpha$and$\beta$  will be gamma distribution also,
$$\pi_2=f(\alpha|x_1,...,x_{r_1})=\frac{(v_1+\sum_{i=1}^{r_1}x_i+x_{r_1}(n-r_1))^{u_1+r_1}(\frac{1}{\alpha})^{r_1+u_1}e^{-\frac{(\sum_{i=1}^{r_1}x_i+x_{r_1}(n-r_1))}{\alpha}}}{\Gamma(r_1+u_1+1)},\eqno(3.9)$$
and
$$\pi_3=f(\beta|y_1,...,y_{r_2})=\frac{(v_2+\sum_{j=1}^{r_2}y_j+y_{r_2}(m-r_2))^{u_2+r_2}(\frac{1}{\beta})^{r_2+u_2}e^{-\frac{(\sum_{j=1}^{r_2}y_j+y_{r_2}(m-r_2))}{\beta}}}{\Gamma(r_2+u_2+1)},\eqno(3.10)$$
the joint posterior function, put,\\
$$\zeta=(v_1+\sum_{i=1}^{r_1}x_i+x_{r_1}(n-r_1)),\tau=(v_2+\sum_{j=1}^{r_2}y_j+y_{r_2}(m-r_2))$$and
$~~~~~~~~~~~~~~~~~~~~~~~~~~~~~~~~~~~~~~k=\frac{1}{\Gamma(r_1+u_1+1)\Gamma(r_2+u_2+1)}$
$$\pi(\alpha,\beta|x,y)=k\zeta^{(u_1+r_1)}\tau^{(u_2+r_2)}(\frac{1}{\alpha})^{r_1+u_1}(\frac{1}{\beta})^{r_2+u_2}e^{\frac{-\zeta}{\alpha}}e^{\frac{-\tau}{\beta}},\eqno(3.11)$$
Hence Bayes estimator $\hat{R_3}$ of $R$  will be
$$\hat{R_3}=E(R|x,y)=\frac{k\zeta^{(u_1+r_1)}\tau^{(u_2+r_2)}}{\Gamma(r_1+u_1+r_2+u_2+1)}\int_{0}^{1}\frac{R^{(r_1+u_1)}(1-R)^{(r_2+u_2+1)}}{((1-R)\zeta+R\tau)^{u_1+r_1+u_2+r_2}}dR,\eqno(3.12)$$
From equation (3.12) there is no explicit form of   $\hat{R_3}$so, The computation of the Bayes estimator $\hat{R_3}$,is very complicated ,we will use the MATHCAD program to evaluate the value of $\hat{R_3}$ .
\subsection*{3.3.2. Non Informative Prior Distributions}
Let $X_1,...,X_{r_1}$ be a random sample from exponential distribution with parameter $\alpha$ . The prior distribution of $\alpha$  is proportional to $\sqrt{I(\alpha)}$ , where $I(\alpha)$   is Fisher's information of the sample about $\alpha$ , and is given by
$$I(\alpha)=\frac{1}{\alpha^2},\eqno(3.13)$$
from that the prior distribution
$$\pi_1\propto\frac{1}{\alpha},\eqno(3.14)$$
Similarly, if  $Y_1,...,Y_{r_2}$is a random sample from exponential distribution with parameter $\beta$, the prior distribution of  $\beta$ will be given by:
$$\pi_2\propto\frac{1}{\beta},\eqno(3.15)$$
if we have  $\alpha$  and  $\beta$ are independent then the posterior joint distribution of   $\alpha$  and  $\beta$ ,will be
$$\pi(\alpha,\beta|x_1,...,x_{r_1},y_1,...,y_{r_2})\propto{L(x_1,...,x_{r_1}|\alpha)L(y_1,...,y_{r_2}|\beta)\pi_1(\alpha)\pi_2(\beta)},\eqno(3.16)$$
then
$$\pi(\alpha,\beta|x_1,...,x_{r_1},y_1,...,y_{r_2})=\frac{n!}{(n-r_1)!}(\frac{1}{\alpha})^{r_1}e^{-\frac{1}{\alpha}(\sum_{i=1}{r_1}x_i+x_{r_1}(n-r_1))}\frac{m!}{(m-r_2)!}$$$$(\frac{1}{\beta})^{r_2}e^{-\frac{1}{\beta}(\sum_{j=1}{r_2}y_j+y_{r_2}(m-r_2))},\alpha,\beta>0$$
put
$$\delta=(\sum_{i=1}{r_1}x_i+x_{r_1}(n-r_1)),\varepsilon=(\sum_{j=1}{r_2}y_j+y_{r_2}(m-r_2))$$
$$\pi(\alpha,\beta|x_1,...,x_{r_1},y_1,...,y_{r_2})=\frac{n!m!\alpha^{-(1+r_1)}\beta^{-(1+r_2)}e^{-\frac{\delta}{\alpha}}e^{-\frac{\varepsilon}{\beta}}}{(n-r_1)!(m-r_2)!},\alpha,\beta>0,\eqno(3.17)$$
under the mean square error, Bayes estimator $\hat{R_4}$ of $R$ will be
$$\hat{R_4}=E(R|x,y)=\frac{n!m!}{(n-r_1)!(m-r_2)!(r_1+r_2+3)!}\int_{0}^{1}\frac{R^{(r_2+2)}(1-R)^{(r_1+1)}}{((1-R)\delta+R\varepsilon)^{r_1+r_2+2}}dR,\eqno(3.18)$$
From equation (3.18) there is no explicit form of   $\hat{R_4}$so, The computation of the Bayes estimator $\hat{R_4}$,is very complicated ,we will use the MATHCAD program to evaluate the value of $\hat{R_4}$ .

\section*{4.~Interval Estimation of $R$ }
\subsection*{4.1~   Approximate confidence interval }
Let  $X_1,...,X_{r_1}$  and  $Y_1,...,Y_{r_2}$ be a random samples with size $r_1,r_2$ from exponential distribution with parameter  $\alpha,\beta$ respectively ,we can show that the maximum likelihood function of $R$ ,$\hat{R}$  is asymptotically normal distribution with mean $R$ and variance-covariance matrices  $I^{-1}(\alpha,\beta)$ where $I(\alpha,\beta)$ is the Fisher information matrix and given by:
$$~I(\alpha,\beta)=
\pmatrix{\frac{r_1}{\alpha^2}&0\cr
0&\frac{r_2}{\beta^2}\cr}$$

The variance-covariance matrix is obtained by inverting the information matrix with elements that are negatives of the expected values of the second order derivatives of logarithms of the likelihood functions, and the asymptotic variance-covariance matrix is obtained by replacing values by their maximum likelihood estimators. Hence, the asymptotic variance-covariance matrix will be
$$~I^{-1}(\alpha,\beta)=
\pmatrix{\frac{\beta^2}{r_2}&0\cr
0&\frac{\alpha^2}{r_1}\cr}$$
From (Surles and Padjett (2001)) that the maximum likelihood function of $R$ ,$\hat{R_1}$   is asymptotically normal distribution with mean $R$  and variance

$$\sigma_{\hat{R_1}}^2=\frac{r_1r_2(\alpha+\beta)^2}{\alpha^2\beta^2}$$
Hence,$(1-\alpha)$100 an approximate confidence interval for $R$ would be $(L_1,U_1)$ ,and the value of $L_1,U_1$ is given as:
$$L_1=\hat{R_1}-z_{1-\frac{\alpha}{2}}\sigma_{\hat{R_1}},\eqno(4.1)$$
and
$$U_1=\hat{R_1}+z_{1-\frac{\alpha}{2}}\sigma_{\hat{R_1}},\eqno(4.2)$$
where $z_{1-\frac{\alpha}{2}}$  quantile of the standard normal distribution and $\hat{R_1}$ is given by Eq.(3.1).
\subsection*{4.2~   Exact confidence interval }

Let  $X_1,...,X_{r_1}$ and  $Y_1,...,Y_{r_2}$ be random samples with size $r_1,r_2$ from exponential distribution with parameter $\alpha,\beta$ respectively.\\
Since $z_i=lne^{x_i}$and $v_j=lne^{y_j}$  are independent with gamma distribution with parameters $(r_1,\alpha)$ and $(r_2,\beta)$ respectively,$2\alpha{lne^{\sum_{i=1}^{r_1}x_i}}$  and $2\beta{lne^{\sum_{j=1}^{r_2}y_j}}$ are independent with chi square distributions with degree of freedom $2r_1$ and $2r_2$ hence
$\hat{rR_1}=(1+\frac{\beta}{\alpha})^{-1}$\\
we know that $F_1=frac{V\alpha}{Z\beta}$ has F-distribution with $(2r_1,2r_2)$degrees of freedom ,$\hat{rR_1}=(1+\frac{\beta}{\alpha}F_1)^{-1}$ ,this equation can be written as follow $F_1=\frac{(1-\hat{R_1})}{\hat{R_1}}\frac{R}{(1-R)}$\\
using $~F_1~$as a pivotal quantity,we obtain a $~(1-\alpha)~$100$\%$~~confidence interval for $~R~$as
$$~L_2=F_{1-\frac{\alpha}{2}}(2r_2,2r_1)(F_{1-\frac{\alpha}{2}}(2r_2,2r_1)+\frac{V}{Z})^{-1}~$$
and
$$~U_2=F_{\frac{\alpha}{2}}(2r_2,2r_1)(F_{\frac{\alpha}{2}}(2r_2,2r_1)+\frac{V}{Z})^{-1}~$$
where\\
$F_{\alpha}$is $(1-\alpha)$th of an F distribution random variables with $~(2r_2,2r_1)~$degrees of freedom.

\section*{5.~Numerical illustrations}
in this section we will compare the different point estimators of $~R~$ , namely $~{\hat{R_{1}}}$,${\hat{R_{2}}}$,${\hat{R_{3}}}~$and$~{\hat{R_{4}}}~$and in the cases of $~R=0.25,R=0.4,R=0.5~$and$~R=0.538~$,and different values of their parameters $\alpha$ ,$\beta$ . 2999 samples are generated of various size of  $~n=5,10,15,20~$, $~n=25~$and $~n=50~$ from exponential distribution with parameters $\alpha$ ,$\beta$ respectively.

\subsection*{5.1.~Numerical illustrations in the case $~r_1=r_2~$ in tables(1-4)}
1-MLE has the smallest mean square errors expect at some points of the UMVUE is the smallest. and for $\alpha=2,\beta=6~$.both Bayes estimator and  Non-Informative estimator are the smallest.\\
2-For some values of $n,m,r_1,r_2,\alpha$ and $\beta$ Bayesian estimator has advantage over UMVUE comparing the mean square errors, such that:\\
$*$ In table (1) in case $\alpha=2,\beta=3~$at $n=m=5,r_1=r_2=3$.\\
$*$ In table (2) in case $\alpha=2,\beta=6~$  for all values of $n,m$ except at $n=m=5,r_1=r_2=4$.\\
$*$ In table (4) in case $\alpha=\beta=7~$ for the values of $n=m=15,r_1=r_2=12$ .\\
3-For all values of $n,m,r_1$ and $r_2$ Bayesian estimator equal to Non-Informative estimator.
\subsection*{5.2.~Numerical illustrations in the case $~r_1\not=r_2~$ in tables(5-7)}
1-MLE has the smallest mean square errors expect at some points of the UMVUE is the smallest .and for $\alpha=2,\beta=6~$ .both Bayes estimator and  Non-Informative estimator are the smallest.\\
2-For some values of $n,m,r_1,r_2,\alpha$ and $\beta$ UMVUE has advantage over MLE comparing the mean square errors, such that:\\
$*$In table (5) in case $\alpha=2,\beta=3~$ for the values of  $n=5,m=4,r_1=3,r_2=2$ and in the case $n=10,m=5,r_1=6,r_2=5$, also in the case $n=25,m=10,r_1=20,r_2=9$ . \\
$*$In table (6) in case $\alpha=2,\beta=6~$ the UMVUE is the smallest in the all cases expect for the values of $n=10,m=25,r_1=10,r_2=20$ and in the case $n=15,m=25,r_1=10,r_2=20$  ,also in the case and $n=50,m=25,r_1=4,r_2=24$.\\
$*$In table (7) in case $\alpha=\beta=7~$ the UMVUE is the smallest in the all cases expect  for the values of $n=5,m=4,r_1=3,r_2=2$ and $n=5,m=5,r_1=4,r_2=3$.\\
3-For some values of $n,m,r_1,r_2,\alpha$ and $\beta$ Bayesian estimator is equal to UMVUE comparing the mean square errors.\\
4-all mean square errors (MSE1, MSE2, MSE3 gamma and MSE4) increases as $\alpha$ or $\beta$ increases\\
when we compare the case of $~r_1=r_2~$ and $~r_1\not=r_2~$ we have that, in case $~r_1\not=r_2~$ the mean square errors of all points estimators are smaller than the mean square errors in  $~r_1=r_2~$  case, expect some points.
\pagebreak
\subsection*{Comparisons between different estimators for $R=P(Y<X)$ in censored exponential case}
$$Table(1)$$
\begin{tabular}{|c|r|r|c|r|c|r|c|r|c|r|c|r|c|r|c|r|c|r|c|r|c|r|c|r||}
\hline\hline
$m$ & $n$ & $r_1$ & $r_2$ &$R_1$&MSE1&$R_2$&MSE2&$R_3$&MSE3&$R_4$&MSE4 \\\hline\hline
5&5&3&3&0.276&0.0150&0.536&1.39&0.00177&0.159&0.0000432&0.160\\\hline
5&5&4&4&0.631&0.0530&1.90&0.702&0.00&0.160&0.00&0.160\\\hline
10&10&6&6&0.62&0.05&0.66&0.25&0.00& 0.16&   0.00&   0.16\\\hline
10&10&7&7&0.70&0.09&2.14&0.00&0.00& 0.16&   0.00&   0.16\\\hline
10&10&8&8&0.44&0.00&0.58&0.02&0.00& 0.16&   0.00&0.16\\\hline
10&10&9&9&0.76&0.13&0.63&0.68&0.00&0.16&0.00&0.16\\\hline
15&15&12&12&0.70&   0.09&   0.73&   0.59&   0.00&   0.16    &0.00   &0.16\\\hline
15&15&13&13&0.59&0.04&0.58&0.52&0.00&0.16&0.00&0.16\\\hline
15&15&14&14&0.53&   0.02&   0.59    &0.57   &0.00&0.16  &0.00&0.16\\\hline
20&20&15&15&0.66&   0.07    &0.56   &0.53   &0.00   &0.16&0.00& 0.16\\\hline
20&20&16&16&0.54&   0.02&   0.64&   0.75    &0.00&0.16& 0.00    &0.16\\\hline
20&20&17&17&0.73&   0.11&   0.70    &0.00&0.00& 0.16&   0.00    &0.16\\\hline
25&25&23&23&0.60&   0.04&   0.00&   0.61&0.00&0.16&0.00&0.16\\\hline
25&25&24&24&0.48    &0.01   &0.67&0.73&0.00&0.16&0.00&0.16\\\hline
50&50&4&4&0.60&0.04&0.73&0.73&0.00&0.16&0.00&0.16\\\hline
50&50&6&6&0.37&0.00&0.68&0.76   &0.00   &0.16   &0.00&0.16\\\hline
50&50&9&9&0.67&0.08&0.61&0.45   &0.00   &0.16&0.00& 0.16\\\hline
\end{tabular}
$$\alpha=2,\beta=3andR=0.4$$
\pagebreak
$$Table(2)$$
\begin{tabular}{|c|r|r|c|r|c|r|c|r|c|r|c|r|c|r|c|r|c|r|c|r|c|r|c|r||}
\hline\hline
$m$ & $n$ & $r_1$ & $r_2$ &$R_1$&MSE1&$R_2$&MSE2&$R_3$&MSE3&$R_4$&MSE4 \\\hline\hline
5&5&3&3&0.725&0.226&0.526&0.076&0.0003457&0.062 &0.051&0.04\\\hline
5&5&4&4&0.789&0.291&2.244&3.978&8.594E-11&0.062 &0.0006678& 0.062\\\hline
10&10&6&6&0.773&0.273&0.668&0.175&0&0.063&0&0.063\\\hline
10&10&7&7&0.657&0.165&0.789&0.29&0&0.063&0&0.063\\\hline
10&10&8&8&0.671&0.177&0.656&0.164&0&0.063&0&0.063\\\hline
10&10&9&9&0.708&0.209&0.619&0.136&0&0.063&0&0.063\\\hline
15&15&12&12&0.657&0.165&0.645&0.156&0&0.063&0&0.063\\\hline
15&15&13&13&0.741&0.241&0.783&0.284&0&0.063&0&0.063\\\hline
15&15&14&14&0.812&0.316&0.673&0.179&0&0.063&0&0.063\\\hline
20&20&15&15&0.662&0.17&0.779&0.28&0&0.063&0&0.063\\\hline
20&20&16&16&0.71&0.212&0.679&0.184&0&0.063&0&0.063\\\hline
20&20&17&17&0.727&0.228&0.719&0.22&0&0.063&0&0.063\\\hline
25&25&23&23&0.799   &0.301&0.734&0.234&0&0.063&0&   0.063\\\hline
25&25&24&24&0.643   &0.154&0.78 &0.281&0&0.063&0&   0.063\\\hline
50&50&4&4&0.518&0.072&0.771&0.272&0&0.063&0&0.063\\\hline
50&50&6&6&0.688&0.192&0.571&0.103&0&0.063&0&0.063\\\hline
50&50&9&9&0.794&0.296&0.734&0.234&0&0.063&0&0.063\\\hline
\end{tabular}
$$\alpha=2,\beta=6andR=0.25$$
\pagebreak
$$Table(3)$$
\begin{tabular}{|c|r|r|c|r|c|r|c|r|c|r|c|r|c|r|c|r|c|r|c|r|c|r|c|r||}
\hline\hline
$m$ & $n$ & $r_1$ & $r_2$ &$R_1$&MSE1&$R_2$&MSE2&$R_3$&MSE3&$R_4$&MSE4 \\\hline\hline
5&5&    3&3&    0.486&0.002721&0.896&0.128&0.0029&0.287&0.000016&0.29\\\hline
5&5&4&4&0.546&0.00006&0.91&0.138&6.109E-08&0.29&1.652&1.239\\\hline
10&10&6&6&0.516&0.0005&0.532&0.000045&0&0.29&0&0.29\\\hline
10&10&7&7&0.607&0.0046&0.639&0.01&0&0.29&0&0.29\\\hline
10&10&8&8&0.563&0.00059&0.569&0.00095&0&0.29&0&0.29\\\hline
10&10&9&9&0.556&0.00031&1.703&1.357&0&0.29&0&0.29\\\hline
15&15&12&12&0.434&0.011&0.578&0.0016&0&0.29&0&0.29\\\hline
15&15&13&13&0.353&0.035&0.545&0.00004&0&0.29&0& 0.29\\\hline
15&15&14&14&0.413   &0.016&0.616&0.006&0&0.29&0&0.29\\\hline
20&20&15&15 &0.429&0.012&0.532&0.0000417&   0&0.29&0&0.29\\\hline
20&20&16&16&0.543   &0.000023&0.00045&0.289 &0&0.29&0&  0.29\\\hline
20&20&17&17&0.482   &0.0032&0.639&0.01&0&   0.29&0&0.29\\\hline
25&25   &23&23&0.526&0.00016&0.0000053&0.29&0&0.29&0&   0.29\\\hline
25&25&24&24&0.46&   0.006215&4.86E-12&0.29&0&0.29   &0&0.29\\\hline
50&50&4&4&0.484&0.002912&0.663&0.016&0&0.29&0&0.29\\\hline
50&50   &6&6&0.443&0.009077&0.72&0.033&0&0.29&0&0.29\\\hline
50&50&9&9&0.453&0.007321&0.412&0.016&0&0.29&0&0.29\\\hline
\end{tabular}
$$\alpha=7,\beta=6andR=0.538$$
\pagebreak
$$Table(4)$$
\begin{tabular}{|c|r|r|c|r|c|r|c|r|c|r|c|r|c|r|c|r|c|r|c|r|c|r|c|r||}
\hline\hline
$m$ & $n$ & $r_1$ & $r_2$ &$R_1$&MSE1&$R_2$&MSE2&$R_3$&MSE3&$R_4$&MSE4 \\\hline\hline
5&5&3&3&0.535&0.001238&0.549&0.002414&0.007157&0.243&0.000089&0.25\\\hline
5&5&4&4&0.286&0.046&0.592&0.008417&0.29&0.044&0.0000024&0.25\\\hline
10&10&6&6&0.681&0.033&0.562&0.003845&0.00085&0.249&0&0.25\\\hline
10&10&7&7&0.519&0.00035&0.7&0.04&0& 0.25&0&0.25\\\hline
10&10   &8&8&0.426&0.00546&0.713&0.046&0&0.25&0&0.25\\\hline
10&10&9&9&0.412&0.007753&0.628&0.017&0&0.25&0&0.25\\\hline
15&15&12&12&0.525&0.0006029&1.619&1.252&0&0.25  &0& 0.25\\\hline
15&15&13&13&0.628&0.016&0.055&0.198&0&0.25&0&0.25\\\hline
15&15&14&14&0.682&0.033&0.6&0.01&0&0.25&0&0.25\\\hline
20&20&15&15&0.43&0.00485&0.687&0.035&0&0.25&0&0.25\\\hline
20&20&16&16&0.487&0.0001711&0.004066&0.246&0&0.25&0&    0.25\\\hline
20&20&17&17&0.426&0.00552&0.000001057&0.25&0&0.25&0&    0.25\\\hline
25&25&23&23&0.366&0.018&0.591&0.008315&0&0.25&0&0.25\\\hline
25&25&24&24&0.46&0.006215&8.178E-14&0.25&0&0.25&0&0.25\\\hline
50&50&4&4&0.516&0.0002437&0.647&0.022&0&0.25&0&0.25\\\hline
50&50&6&6&0.487&0.0001691&0.659&0.025&0&0.25&0&0.25\\\hline
50&50&9&9&0.443&0.003232&0.206&0.086&0&0.25&0&0.25\\\hline
\end{tabular}
$$\alpha=7,\beta=7andR=0.5$$
\pagebreak
$$Table(5)$$
\begin{tabular}{|c|r|r|c|r|c|r|c|r|c|r|c|r|c|r|c|r|c|r|c|r|c|r|c|r||}
\hline\hline
$m$ & $n$ & $r_1$ & $r_2$ &$R_1$&MSE1&$R_2$&MSE2&$R_3$&MSE3&$R_4$&MSE4 \\\hline\hline
5&4&3&2&0.677&0.077&0.362&0.001465& 8.027E-09&0.16& 0.00002938& 0.16\\\hline
5&5&4&3&0.521&0.015&0.759&0.129&2.095E-11   &0.16   &6.352E-08  &0.16\\\hline
10&5&   6&5&0.78&0.144&0.457&0.003292   &0&0.16&0&  0.16\\\hline
10&10&7&6&0.56&0.026&0.716&0.1&0&0.16&0&0.16\\\hline
10&15&8&7&0.638&    0.057   &0.62&0.049 &0&0.16&0&0.16\\\hline
10&20&9&10& 0.611&0.044&0.0001796&  0.16&   0&0.16&0&0.16\\\hline
10&25&10&20&0.524   &0.015&0.000001812&0.16&0&0.16&0&0.16\\\hline
15&5&6&5&0.796&0.157&0.567&0.028&0&0.16&0   &0.16\\\hline
15&10&7&6&0.533&0.018&0.8&0.16&0&0.16&0&0.16\\\hline
15&15&8&7&0.546&0.021&0.621&0.049&0&0.16&0&0.16\\\hline
15&20&9&10&0.518&0.014&0.601&0.04&0&0.16&0&0.16\\\hline
15&25&10&20&0.589   &0.036&0.001212&0.159&0&0.16&0&0.16\\\hline
20&5&15&4&0.325&0.005696&0.65   &0.063&0&0.16&0&0.16\\\hline
20&10&16&9&0.614&0.046&0.59&0.036&0&0.16&0&0.16\\\hline
20&15&17&14&0.551&0.023&0.531&0.017&0&0.16&0&0.16\\\hline
25&10&20&9&0.724&0.105&0.689&0.083&0&0.16&0&0.16\\\hline
25&15&22&14&0.657&0.066&0.644&0.06&0&0.16&0&0.16\\\hline
25&20&24&19&0.569&0.029&0.574   &0.03   &0&0.16&0&0.16\\\hline
50&25&4&24&0.806&0.165&0.226&0.03&0&0.16&0&0.16\\\hline
50&30&6&29&0.727&0.107&0.018&0.146&0&0.16&0&0.16\\\hline
50&40&9&39&0.759&0.129& 0.0001917&0.16&0&0.16&0&0.16\\\hline
\end{tabular}
$$\alpha=2,\beta=3andR=0.4$$
\pagebreak
$$Table(6)$$
\begin{tabular}{|c|r|r|c|r|c|r|c|r|c|r|c|r|c|r|c|r|c|r|c|r|c|r|c|r||}
\hline\hline
$m$ & $n$ & $r_1$ & $r_2$ &$R_1$&MSE1&$R_2$&MSE2&$R_3$&MSE3&$R_4$&MSE4 \\\hline\hline
5&4&3&2&0.911&0.436&0.907&0.432&0.0003433&0.062&0.001571&0.062\\\hline
5&5&4&3&0.715&0.216&0.811&0.314&1.118E-13&0.062 &5.941E-07& 0.062\\\hline
10&5&   6&5&0.805&0.308&0.875&0.39&0&0.062&0&0.063\\\hline
10&10&7&6&0.837&0.344&0.825&0.33&0&0.062&0&0.063\\\hline
10&15&8&7&0.727&0.228&0.739&0.24&0&0.062&   0&0.063\\\hline
10&20&9&10& 0.71&0.212&0.73&0.231&0&0.062&0&0.063\\\hline
10&25&10&20&0.762&0.262 &0.0001145& 0.062   &0&0.062&0&0.063\\\hline
15&5&   6&5&0.865&0.378&0.789&0.291&0&0.062&0&0.063\\\hline
15&10&7&6&0.702&0.204&0.8&0.302&0&0.062&0   &0.063\\\hline
15&15&8&7&0.801&0.304&0.59&0.116&0& 0.062   &0&0.063\\\hline
15&20&9&10& 0.792   &0.294&0.776&0.277&0&0.062&0&0.063\\\hline
15&25&10&20&0.729   &0.23   &0.003433&0.061&0&0.062&0&0.063\\\hline
20&5&   15&4&0.583& 0.111   &0.718&0.219&0&0.062&0&0.063\\\hline
20&10&16&9& 0.587&0.114&0.72&   0.221&0&0.062&0&0.063\\\hline
20&15&17&14&0.781&0.282&0.862   &0.374&0&0.062&0&0.063\\\hline
25&10&20&9& 0.738&0.238&0.758&0.258 &0&0.062&0& 0.063\\\hline
25&15&22&14&0.71&0.211&0.663&0.17&0&0.062&0&0.063\\\hline
25&20&24&19&0.78&0.281& 0.81&   0.314   &0&0.062&0& 0.063\\\hline
50&25&4&24&0.841&0.35&0.395&0.021&0&0.062&0&0.063\\\hline
50&30&6&29& 0.818   &0.323&0.889&0.408&0&0.062&0&0.063\\\hline
50&40&9&39& 0.859   &0.371&0.00000394&0.062 &0&0.062&0&0.063\\\hline
\end{tabular}
$$\alpha=2,\beta=6andR=0.25$$
\pagebreak
$$Table(7)$$
\begin{tabular}{|c|r|r|c|r|c|r|c|r|c|r|c|r|c|r|c|r|c|r|c|r|c|r|c|r||}
\hline\hline
$m$ & $n$ & $r_1$ & $r_2$ &$R_1$&MSE1&$R_2$&MSE2&$R_3$&MSE3&$R_4$&MSE4 \\\hline\hline
5&4&3&2&0.61&0.012&0.436&0.004035&0.013&0.237&0.022&0.229\\\hline
5&5&4&3&0.259&0.058&0.509&0.00007698&0.000102&0.25&0.009586 &0.241\\\hline
10&5&   6&5&0.427&0.005336&0.602&0.01&0.0006316&0.249&0&0.25\\\hline
10&10&7&6&0.381&0.014&0.66&0.026&0& 0.25&0&0.25\\\hline
10&15&8&7&0.389&0.012&0.745&0.06&0& 0.25&0&0.25\\\hline
10&20&9&10& 0.012   &0.001402&0.369&0.017&0&0.25&   0&0.25\\\hline
10&25&10&20&0.363&0.019 &0.06&0.194 &0&0.25&0&0.25\\\hline
15&5&   6&5&0.439&0.003772&0.728&0.052&0&0.25&0&0.25\\\hline
15&10&7&6&0.723&0.05&0.715&0.046&0&0.25&0&0.25\\\hline
15&15&8&7&0.535&0.001225&0.608&0.012&0&0.25&0&0.25\\\hline
15&20&9&10& 0.66&0.025&0.628&0.016&0&0.25&0&0.25\\\hline
15&25&10&20&0.639   &0.019&0.178&0.104&0&0.25&0&0.25\\\hline
20&5&15&4&0.505&0.00002566&0.406&0.008777&0&0.25&0&0.25\\\hline
20&10&16&9& 0.541   &0.001673&0.609&0.012&0 &0.25   &0&0.25\\\hline
20&15&17&14&0.472&0.0007671&0.00226&0.248   &0&0.25&0&0.25\\\hline
25&10&20&9&0.487&0.0001627&0.551&0.002556&0&0.25&0&0.25\\\hline
25&15&22&14&0.509   &0.00008313 &0.487&0.0001641&0&0.25&0&0.25\\\hline
25&20&24&19&0.528   &0.0007997& 0.576   &0.00584&0&0.25&0&0.25\\\hline
50&25&4&24& 0.791&0.085 &1.68   &1.392&0&0.25&0&0.25\\\hline
50&30&6&29& 0.752   &0.063&1.144&0.415&0&0.25&0&0.25\\\hline
50&40&9&39& 0.755   &0.065&0.023&0.227&0&0.25&0&0.25\\\hline
\end{tabular}
$$\alpha=7,\beta=7 and R=0.5$$

Where both  $\alpha$ and $\beta$ are the scale parameters for $X$ and $Y$ respectively. $R$ is the reliability function ,also. MSE1 is The mean square error of $~{\hat{R_{1}}}~$ , MSE2 is The mean square error of $~{\hat{R_{2}}}~$, MSE3 is The mean square error of $~{\hat{R_{3}}}~$ and MSE4 is The mean square error of $~{\hat{R_{4}}}~$.

\begin{center}
{\large{\bf {References}}}
\end{center}
\quad\quad\enskip~~~Awad, A. M. , Hamdan, M. A. and Azzam, M. M.(1981). Some inference results on  $Pr{(Y<X)}$ in the bivariate exponential model. Commun. Statist. Theory and Methods, A10(24),2515-2525.\\

\quad\quad\enskip~~~Ayman , B. and Walid, A.(2003). Shrinkage Estimation of $P(Y<X)$ in the exponential case. Commun. Statist. Simula, 32(1), 31-42.\\

\quad\quad\enskip~~~Bai, D. S. and Hong, Y. W. (1992). Estimation of $P(Y<X)$  in the exponential  case with common location parameter. Communication Statistics: Theory and Methods, 21(1), 269-282.\\

\quad\quad\enskip~~~Basu, A. (1981). The estimation $P(X>Y)$ for distributions useful in life testing. Naval Research Logistics Quarterly, 3,383-392.\\

\quad\quad\enskip~~~Beg, M. A. (1980a). Estimation of $Pr{Y < X}$ for exponential family. IEEE Trans of reliability, Vol. R-29, 158-159.\\

\quad\quad\enskip~~~Beg, M. A. (1980b). On estimation of $Pr{Y < X}$ for two-parameter exponential ". Metrika, 27,29-34.\\

\quad\quad\enskip~~~Beg, M. A. (1980c). Estimation of $Pr{Y < X}$ for truncation parameter distribution . Communication in Statistics: Theory and methods,9(3),327-345.\\

\quad\quad\enskip~~~Chao, A. (1982): "On comparing estimators of $Pr{Y < X}$ in the exponential case", IEEE trans. On Reliability, Vol. 31, 389-392.\\

\quad\quad\enskip~~~Chiou, W.J. and  Cohen, A.(1984).estimating the common location  parameter of exponential distrinution with censored sample, Naval Research Logistic Quarterly 31, 475-482.\\

\quad\quad\enskip~~~Cramer, E. and Kamps, U. (1997). A note on the UMVUE of $Pr{Y < X}$in the exponential case. Communication Statistics: Theory and Methods, 26(4),1051-1055.\\

\quad\quad\enskip~~~Dinh, K. T. Singh, J. and Gupta, R. C. (1991). Estimation of reliability  in bivariate distributions. Statistics,22(3),409-417.\\

\quad\quad\enskip~~~Ghogh, M. and Razmpour, A.(1984). Estimation of the common  location parameter of several exponentials, Sankhya  A46,383-394.\\

\quad\quad\enskip~~~Jeevanand, E. S. and nair, N. U. (1994) Estimating $P(X>Y)$ from  exponential samples containing spurious observations. Communication Statistics: Theory and Methods, 23(9), 2629-2642.\\

\quad\quad\enskip~~~Johnson, R. A. (1988). Stress- Strength Models for Reliability. Handbook of Statistics , 7 (P.R. Kirshnaiah and C.R. Rao, ed.) , Amsterdam,  North Holland.\\

\quad\quad\enskip~~~Khayar A. (2001). A study on reliability of sress-strength models. Ph. D. Thesis, Faculty of Science, Azhar University.\\

\quad\quad\enskip~~~Kunchur, S. H. and Mounoli, S. s. (1993). Estimation of reliability for a multi-component survival stress-strength model based on exponential distributions. Commun. Statist. Simula.,23(1)839-943.\\

\quad\quad\enskip~~~Mood, A. M., Graybill, F. A. and Boes, D. C. (1974). Introduction to the theory of statistics. Third edition, Mc Graw. Hill.\\

\quad\quad\enskip~~~Sathe, Y. S. and Shah, S. P. (1981) "On estimating $Pr{Y < X}$ for the exponential distribution" Commun. Statist. A, Vol. A10 (1), 39-47.\\

\quad\quad\enskip~~~Selvavel, K. , Nanthekumar, A. and Michalek, J. ( 2000).On the estimation $P[X<Y]$ for truncated bivariate exponential distribution and its application .Journal of applied statistical science. Volume 10, No.1 pp. 47-56.\\

\quad\quad\enskip~~~Siu-Keung, T. and Geoffrey, T. (1996).Shrinkage estimation of reliability for exponential distributed lifetimes,  Commun. Statist. Simula. 25(2)415-430.\\

\quad\quad\enskip~~~Surles, J. G. and Padgett, W. J. (2001). Inference for $P(Y<X)$ in the Burr type X model. Journal of Applied Statistical Science, vol. 7, no.4, 225-238.\\

\quad\quad\enskip~~~Tachen, L. (2005). On empirical bayes testing for reliability, Communication Statistics: Theory and Methods,  34,660-670.\\

\quad\quad\enskip~~~Tong, H. (1974) "A note on the estimation of $Pr{Y < X}$ in the exponential case" .Techno metrics. Vol. 16, 625.\\

\quad\quad\enskip~~~Tong, H. (1977) " On the estimation of $Pr[Y< X]$ for exponential families". IEEE Trans. on. Reliability, Vol R-26, No.1, 54-56.\\

\end{document}